\documentclass[12pt]{iopart}

\usepackage{iopams}
\usepackage{graphicx}
\usepackage{hyperref}
\begin{document}

\title{Properties of Light Flavour Baryons in Hypercentral quark model}

\author{Kaushal Thakkar$^*$, Bhavin Patel$^\dag$, Ajay Majethiya$^\ddag$ and P.C.Vinodkumar$^*$}

\address{$^*$Department of Physics, Sardar Patel University, Vallabh Vidyanagar- 388 120, \\
$^\dag$ LDRP, Gandhinagar, Gujarat, $^\ddag$ KITRC, Kalol, Gujarat,
INDIA.

} \ead{kaushal21185@yahoo.co.in}
\begin{abstract}

The light flavour baryons are studied within the quark
model using the hyper central description of the three-body system.
The confinement potential is assumed as hypercentral coulomb plus
power potential ($hCPP_\nu$) with power index $\nu$. The masses and
magnetic moments of light flavour baryons are computed for different
power index, $\nu$ starting from 0.5 to 1.5. The predicted masses
and magnetic moments are found to attain a saturated value with
respect to variation in $\nu$ beyond the power index  $\nu>$ 1.0.
Further we computed transition magnetic moments and radiative decay
width of light flavour baryons. The results are in good agreement
with known experimental as well as other theoretical models.
\end{abstract}

\maketitle

  \def\be{\begin{equation}}      \def\ol{\overline}
  \def\ee{\end{equation}}    \def\beq{\begin{eqnarray}}
  \def\dis{\displaystyle}    \def\eeq{\end{eqnarray}}
  \def\btd{\bigtriangledown}     \def\m{\multicolumn}
  \def\dfac{\dis\frac}       \def\ra{\rightarrow}

\section{Introduction}Baryons are not only the interesting systems to study the quark
dynamics
  and their properties but are also interesting from the point of view of simple systems to
  study three body interactions. In the last two decades, there has been great advancement
  in the study of baryon properties. The ground state masses and magnetic moments of many low
  lying baryons have been measured experimentally. The magnetic
  moments of all octet baryons $(J^P=\frac{1}{2}^+)$ are known
  accurately except for $\Sigma^0$ which has a life time too short. For
  the decuplet baryons $(J^P=\frac{3}{2}^+)$, the experimental
  measurements are poor as they have very short
  life times due to available strong interaction decay channels. The
  $\Omega^-$ is an exception as it is composed of three s
  quarks which decays via
  weak interaction causing longer life time for it \cite{PDG2008}.
   The $\Delta$ particles are produced in scattering the pion,
   photon, or electron beams off a nucleon target. High precision
   measurements of the N $\rightarrow$ $\Delta$ transition by means
   of electromagnetic probes became possible with the advent of the
   new generation of electron beam facilities such as LEGS, BATES,
   ELSA, MAMI, and those at the Jefferson Lab. Many such experimental programs
   devoted to the study of electromagnetic properties of the
   $\Delta$ have been reported in the past few years \cite{M. Kotulla,A. Bosshard 1991,W. M. Yao}.
 The electromagnetic transition of $\Delta \rightarrow
  N\gamma$ have been the subject of intense study
  \cite{D. B. Leinweber,E. Kaxiras,S. Capstick,M. N. Butler,T. M. Aliev 2006,T. M. Aliev }.
The experimental information provides new incentives for
  theoretical study of these observables.\\

  Theoretically, there exist serious discrepancies between the quark model and experimental
  results particularly in the  predictions of their magnetic moments \cite{M.Bagchi 2006,T. M. Aliev 2000,Frank X. Lee 1998}.
  Prediction of transition magnetic moments between the decuplet to octet
  ($\frac{3}{2}^+ \rightarrow \frac{1}{2}^+ $) is as important as the prediction of the masses and magnetic moments
  of the baryons (octet and decuplet) for testing of any model hypothesis
  and understanding the dynamics of quarks and meaning of the constituent mass of the quarks in the hadronic
  scale.
   Various attempts including lattice QCD (Latt) \cite{D. B. Leinweber 1992,I. C. Cloet 2003,{I. C. Cloet}},
  chiral perturbation theory ($\chi$PT) \cite{L. S. Geng,M. N. Butler 1994,Meissner 1997,P. Ha 1998,{S. J. Puglia 2000}}, relativistic
  quark model (RQM) \cite{F. Schlumpf 1993,K. T. Chao 1990}, non relativistic quark
  model (NRQM) \cite{Ha P. and Durand L. 1990}, QCD sum rules (QCDSR) \cite{T. M. Aliev 2000,Frank X. Lee 1998,Frank X. 1998,{Lai Wang 2008}}
  , chiral quark soliton  model ($\chi$QSM)  \cite{H. C. Kim 1998,{H. C. Kim
  2004}}, chiral constituent quark model ($\chi$CQM) \cite{H Dahiya
  2009},   chiral bag model ($\chi$B) \cite{S. T. Hong 1994}, cloudy bag model \cite{M. I. Krivoruchenko 1987},
   quenched lattice gauge theory
  \cite{Leinweber D. B. 1992} etc., have been tried, but with partial success.\\

  The importance of three body interaction in the description of baryon was felt in many
  cases. In this context it is found that
   the six dimensional hyper central model with coulomb plus power
  potential ($hCPP_\nu$)
  is successful in predicting the masses and magnetic moments
  of heavy flavour baryons
  (baryon containing charm or beauty quarks) \cite{Bhavin 2008,Bhavin J
  2008}. Unlike in the case of many other potential models, in
  the $hCPP_\nu$ model, the confinement potential expressed in
  terms of the three body hyper spherical co-ordinate is able to account for the three body
  effects.\\

  Accordingly, in this paper we extend the $hCPP_\nu$ model  in the light flavour baryonic sector
  to compute the masses, magnetic moments of octet and decuplet baryons.
  We also study the electromagnetic transition
 and radiative decay width of those baryons.
  In section 2 the hypercentral scheme and a brief introduction of $hCPP_\nu$ potential employed
   for the present study are described.
  Section 3 describes the computational details of the magnetic moments of octet and
  decuplet baryons and the $\frac{3}{2}^+ \rightarrow \frac{1}{2}^+ $ transition magnetic moments.
  Section 4 describes the radiative decay widths for those transition.
 In section 5, we discuss our results while comparing with other theoretical
  predictions and experimental results and draw important
  conclusions.

\section{Hyper Central Scheme for Baryons}

Quark model description of baryons is a simple three body system of
interest. Generally the phenomenological interactions among the
three quarks are studied using the two-body quark potentials such as
the Isgur Karl Model \cite{Isgur1978}, the Capstick and Isgur
relativistic model \cite{Godfrey1985,Capstick1986}, the Chiral quark
model \cite{H. Dahiya 2003}, the Harmonic Oscillator model
\cite{Murthy1985,Roberts2007} etc. The three-body effects are
incorporated in such models through two-body and three-body
spin-orbit terms \cite{Bhavin 2008,Garcilazo2007}. The Jacobi
Co-ordinates to describe baryon as a bound state of three different
constituent quarks is given by \cite{R. Bijker 2000}

\begin{equation}\label{}
    \vec{\rho}=\frac{1}{\sqrt{2}}(\vec{r}_1-\vec{r}_2)
 \,\,{;}\,\,
    \vec{\lambda}=\frac{(m_1\vec{r}_1+m_2\vec{r}_2-(m_1+m_2)\vec{r}_3)}{\sqrt{m_1^2+m_2^2+(m_1+m_2)^2}}\\
\end{equation}
Such that
\begin{equation}\label{}
m_\rho=\frac{2\,\,m_1\,\, m_2}{m_1+m_2}
 \,\,{;}\,\,
m_\lambda=\frac{2\,m_3\,\,(m_1^2+m_2^2+m_1m_2)}{(m_1+m_2)\,(m_1+m_2+m_3)}
\end{equation}
Here $m_1$, $m_2$ and $m_3$ are the constituent quark mass
parameters.\\

 In the hypercentral model, we introduce the hyper
spherical coordinates which are given by the angles
\begin{equation}\label{}
\Omega_\rho=(\theta_\rho,\phi_\rho)\,\,{;}\,\,
\Omega_\lambda=(\theta_\lambda,\phi_\lambda)
\end{equation}\label{}
together with the hyper radius, $x$ and hyper angle $\xi$ respectively as,\\
\begin{equation}\label{}
x=\sqrt{\rho^2+\lambda^2}\,\,{;}\,\,\xi=\arctan\left(\frac{\rho}{\lambda}\right)\\
\end{equation}\label{}
The model Hamiltonian for baryons can now be expressed as
\begin{equation}\label{eq:404}
H=\frac{P^2_\rho}{2\,m_{\rho}}+\frac{P^2_\lambda}{2\,m_{\lambda}}+\frac{P^2_R}{2\,M}+V(\rho,\lambda)=\frac{P^2_x}{2\,m}+V(x)\\
\end{equation}\label{}
Here the potential $V(x)$ is not purely a two body interaction but
it contains three-body effects also. The three body effects are
desirable in the study of hadrons since the non-abelian nature of
QCD leads to gluon-gluon couplings which produce three-body forces
\cite{Santopinto1998}. Using hyperspherical coordinates, the kinetic
energy operator $\frac{P^2_x}{2\,m}$ of the three-body system can be
written as
\begin{equation}\label{}
\frac{P^2_x}{2\,m}=\frac{-1}{2\,m}\left(\frac{\partial^2}{\partial\,x^2}+\frac{5}{x}\frac{\partial}{\partial\,x}-\frac{L^2(\Omega_\rho,\Omega_\lambda,\xi)}{x^2}\right)\\
\end{equation}\label{}
Where $L^2(\Omega_\rho,\Omega_\lambda,\xi)$ is the quadratic Casimir
operator of the six dimensional rotational group $O(6)$ and its
eigen functions are the hyperspherical harmonics, $Y_{[\gamma]l_\rho
l_\lambda}(\Omega_\rho,\Omega_\lambda,\xi)$ satisfying the
eigenvalue relation
\begin{equation}\label{}
L^2 Y_{[\gamma]l_\rho l_\lambda}(\Omega_\rho,\Omega_\lambda,\xi)=
\gamma(\gamma+4)Y_{[\gamma]l_\rho
l_\lambda}(\Omega_\rho,\Omega_\lambda,\xi)
\end{equation}\label{}
Here $\gamma$ is the grand angular quantum number and it is given by
$\gamma=2\nu+l_\rho+l_\lambda$, and $\nu=0,1,...$ and $l_\rho$ and
$l_\lambda$ being the angular momenta associated with the $\rho$ and
$\lambda$ variables.\\
If the interaction potential is hyper spherical such that the
potential depends only on the hyper radius $x$, then the hyper
radial schrodinger equation corresponds to the hamiltonian given by
Eqn.(\ref{eq:404})  can be written as
\begin{equation}\label{eq:401}
\left[\frac{d^2}{dx^2}+\frac{5}{x}\frac{d}{dx}-\frac{\gamma(\gamma+4)}{x^2}\right]\phi_\gamma(x)=-2m[E-V(x)]\,\phi_\gamma(x)\\
\end{equation}\label{}
where $\gamma$ is the grand angular quantum number.\\

 For the
present study we consider the hyper central potential $V(x)$ as the
hyper coulomb plus power (hCPP$\nu$) form given by \cite{Bhavin
2008,Bhavin J 2008,A 2008}
\begin{equation}\label{eq:410}
V(x)=-\frac{\tau}{x}+\beta x^\nu+\kappa + V_{spin}\\
\end{equation}\label{}
In the above equation the spin independent terms correspond to
confinement potential in the hyperspherical co-ordinates. The form
of the potential though hyper central, belong to a generality of
potentials of the form $-Ar^\alpha+kr^\epsilon+V_{0} $ where
$A,k,\alpha$ and $\epsilon$ are non negative constants where as
$V_{0}$ can have either sign. There are many attempts with different
choices of $\alpha$ and $\epsilon$ to study the hadron properties
\cite{Sameer2006}. For example, Cornell potential has
$\alpha=\epsilon=1$, Lichtenberg potential has
$\alpha=\epsilon=0.75$. Song-Lin potential has $\alpha=\epsilon=0.5$
and the Logarithmic potential of Quigg and Rosner corresponds to
$\alpha=0$,  $\epsilon\rightarrow0$ \cite{Sameer2006}.
 Martin potential corresponds to
$\alpha$=0, $\epsilon=0.1$ \cite{Sameer2006} while Grant, Rosner and
Rynes potential corresponds to $\alpha=0.045$, $\epsilon=0$;
Heikkil\"{a}, T\"{o}rnqusit and Ono potential corresponds to
$\alpha=1$, $\epsilon=2/3$ \cite{Heikkila1984}. Potentials in the
region $0\leq \alpha \leq 1.2$, $0\leq \epsilon \leq 1.1$ of
$\alpha-\epsilon$ values are also been explored \cite{Song1991}. So
it is important to study the behavior of different potential scheme
with different choices of $\alpha$ and $\epsilon$ to know the
dependence of their parameters to the hadron properties. The spin
independent part of potential defined by Eqn.(\ref{eq:410})
corresponds to $\alpha=1$ and $\epsilon=\nu$. Here $\tau$ of the
hyper-coulomb, $\beta$ of the confining term and $\kappa$ are the
model parameters. The parameter $\tau$ is related to the strong
running coupling constant $\alpha_{s}$ as \cite{Bhavin 2008,Bhavin J
2008}
\begin{equation}\label{11.0}
\tau=\frac{2}{3}\,b\,\alpha_s
\end{equation}\label{}
where b is the model parameter, $\frac{2}{3}$ is the color factor
for the baryon. The potential parameters treated here are similar to
the one employed for the study of heavy flavour baryons \cite{Bhavin
2008,Bhavin J 2008}. The strong running coupling constant is
computed using the relation
\begin{equation} \label{}
\alpha_s=\frac{\alpha_s(\mu_0)}{1+\frac{33-2\,n_f}{12\,\pi}\alpha_s(\mu_0)ln(\frac{\mu}{\mu_0})}
\end{equation}\label{}
where $\alpha_s(\mu_{0}=1 GeV)\approx 0.6$ is considered in the
present study. The spin dependent part of the three body interaction
potential of Eqn.(\ref{eq:410}) is taken as \cite{Bhavin
2008,Garcilazo2007}
\begin{equation}\label{20.5}
V_{spin}(x)=-\frac{1}{4}\alpha_s \, \frac{e^\frac{-x}{x_0}} {x
x^{2}_{0}}\sum\limits_{i{<}j}\frac{\vec{\sigma_i} \cdot
\vec{\sigma_j}}{6m_i m_j} \vec{\lambda_i}\cdot \vec{\lambda_j}
\end{equation}\label{15.5}
where, $x_0$ is the hyperfine parameter of the model.\\

 The six
dimensional radial Schrodinger equation described by
Eqn.(\ref{eq:401}) has been solved in the variational scheme with
the hyper coloumb trial radial wave function given by
\cite{Santopinto1998}

\begin{equation}\label{}
\psi_{\omega\gamma}=\left[\frac{(\omega-\gamma)!(2g)^6}{(2\omega+5)(\omega+\gamma+4)!}\right]^\frac{1}{2}(2gx)^\gamma
e^{-gx} L^{2\gamma+4}_{\omega-\gamma}(2gx)
\end{equation}\label{}
The wave function parameter g and hence the  energy eigen value are
obtained by applying virial theorem for a chosen potential index $\nu$.\\

The baryon masses are determined by the sum of the model quark
masses plus kinetic energy, potential energy and the spin  hyperfine
interaction as
\begin{equation}\label{eq:415}
M_B=\sum\limits_{i}m_i+\langle H \rangle
\end{equation}\label{}
For the present calculations, we have employed the same mass
parameters of the light flavour quarks ($m_u$ = 338 MeV, $ m_d$ =
350 MeV, $m_s$ = 500 MeV) as used in \cite{Bhavin 2008}. We fix
other parameters (b of Eqn.(\ref{11.0}) and $x_0$ of
Eqn.(\ref{20.5})) of the model for each choice of $\nu$ using the
experimental center of weight (spin-average) mass and hyper fine
splitting of the octet decuplet baryons. The procedure is repeated
for different choices of $\nu$ and the computed masses of octet and
decuplet baryons are listed in Table \ref{tab:1} and Table
\ref{tab:2} respectively.

\section{ Magnetic moments of light baryons }

Now the magnetic moment of the baryons are computed in terms of its
quarks spin-flavour wave function of the constituent quarks as
\begin{equation}
\mu_B=\sum\limits_{i}\langle\phi_{sf}\mid\mu_{i}
\vec{\sigma}_{i}\mid\phi_{sf}\rangle
\end{equation}
where
\begin{equation}\label{eq:418}
\mu_{i}=\frac{e_{i}}{2m_{i}}
\end{equation}
Here $e_{i}$ and $\sigma_{i}$ represents the charge and the spin of
the quark constituting the baryonic state and $\mid\phi_{sf}\rangle$
represents the spin-flavour wave function of the respective baryonic
state as listed in \cite{Mexicana 2004}. Here, $m_i$ the mass of the
$i^{th}$ quark in the three body baryon is taken as an effective
mass of the constituting quarks as their motions are governed by the
three body force described through the $hCPP_\nu$ potential appeared
in the hamiltonian  \ref{eq:404}. Accordingly, within the baryon the
mass of the quarks may get modified due to its binding interactions
with other two quarks. We account for this bound state effect by
replacing the mass parameter $m_i$ of Eqn.(\ref{eq:418}) by defining
an effective mass to the bound quarks, $m^{eff}_{i}$ as given by
\cite{Bhavin 2008,Bhavin J 2008,{A 2008}}
\begin{equation}\label{eq:417}
m^{eff}_{i}=m_i\left(
1+\frac{\langle H\rangle}{\sum\limits_{i}m_{i}}\right) \\
\end{equation}
such that
 $M_B=\sum\limits_{i=1}^{3}m^{eff}_{i}.$
 The computations are repeated for the different
choices of the flavour combinations of qqq (q = u, d, s).
 The computed magnetic moments of the octet and decuplet baryons are listed in
Table \ref{tab:3} and \ref{tab:4} respectively.\\

\section{Radiative Decay Width}
The radiative decays of baryons provide much better understanding of
the underlying structure of baryons and the dependence on the
constituent quark mass. Though the non-relativistic model of Isgur
and Karl successfully predicted the electromagnetic properties of
the low lying octet baryons but it fails to provide a good
description of the radiative decay of the decuplet baryons
\cite{{Isgur1978},Lang Yu}. Thus, the successful prediction of the
electromagnetic properties of octet baryons as well as the decuplet
baryons become detrimental for all the phenomenological models. The
radiative decay width of the baryons can be computed using the
relation given by \cite{A 2008}
\begin{equation}
\Gamma_{R} = \frac{q^3}{4\pi}  \frac{2}{2J+1}  \frac{e^2}{m_p^2}
 |\mu_{\frac{3}{2}^+ \rightarrow \frac{1}{2}^+}|^2
\end{equation}\label{}
where $m_p$ is the proton mass, $\mu_{\frac{3}{2}^+ \rightarrow
\frac{1}{2}^+}$ is the radiative transition magnetic moments, q is
the photon energy and is given
by $M_{\frac{3}{2}^+} - M_{\frac{1}{2}^+}$.\\

 The transition magnetic
moments for $\frac{3}{2}^+ \rightarrow \frac{1}{2}^+$ are computed
as
\begin{equation}\label{eq:419}
\mu_{\frac{3}{2}^+ \rightarrow \frac{1}{2}^+} =
\sum\limits_{i}\langle\phi_{sf}^ {\frac{3}{2}^+}\mid\mu_{i}
\vec{\sigma}_{i}\mid\phi_{sf}^ {\frac{1}{2}^+}\rangle
\end{equation}
$| \phi_{sf}^ {\frac{3}{2}^+} \rangle$ represent the spin flavour wave
function of the quark composition for the respective decuplet
baryons while $| \phi_{sf}^ {\frac{1}{2}^+} \rangle$ represent the spin
flavour wave
function of the quark composition for the octet baryons.
The value of $\mu_i$ is given by Eqn.(\ref{eq:418}) and the
$m_{i}^{eff}$ for the transition is calculated using geometric mean
of effective quark masses of decuplet and octet baryons as given by
\cite{Rohit Dhir 2009 }
\begin{equation}
m_{i}^{eff} = \sqrt{m^{eff}_{i({\frac{3}{2}}^+)}
m^{eff}_{i({\frac{1}{2}}^+)}}
\end{equation}
The calculated transition magnetic moments  are listed in Table
\ref{tab:5}\\

 We also calculate the
branching ratio $\frac{\Gamma_{R}}{\Gamma(Baryon)}$ using the
experimental total decay width $\Gamma(Baryon)$ of the respective
decuplet baryons. The computed values of radiative decay width and
the branching ratio for different choices of the potential power
indices are listed in Table \ref{tab:6}.
%

\begin{figure}
\begin{center}
\includegraphics[height=3.5in,width=3.75in]{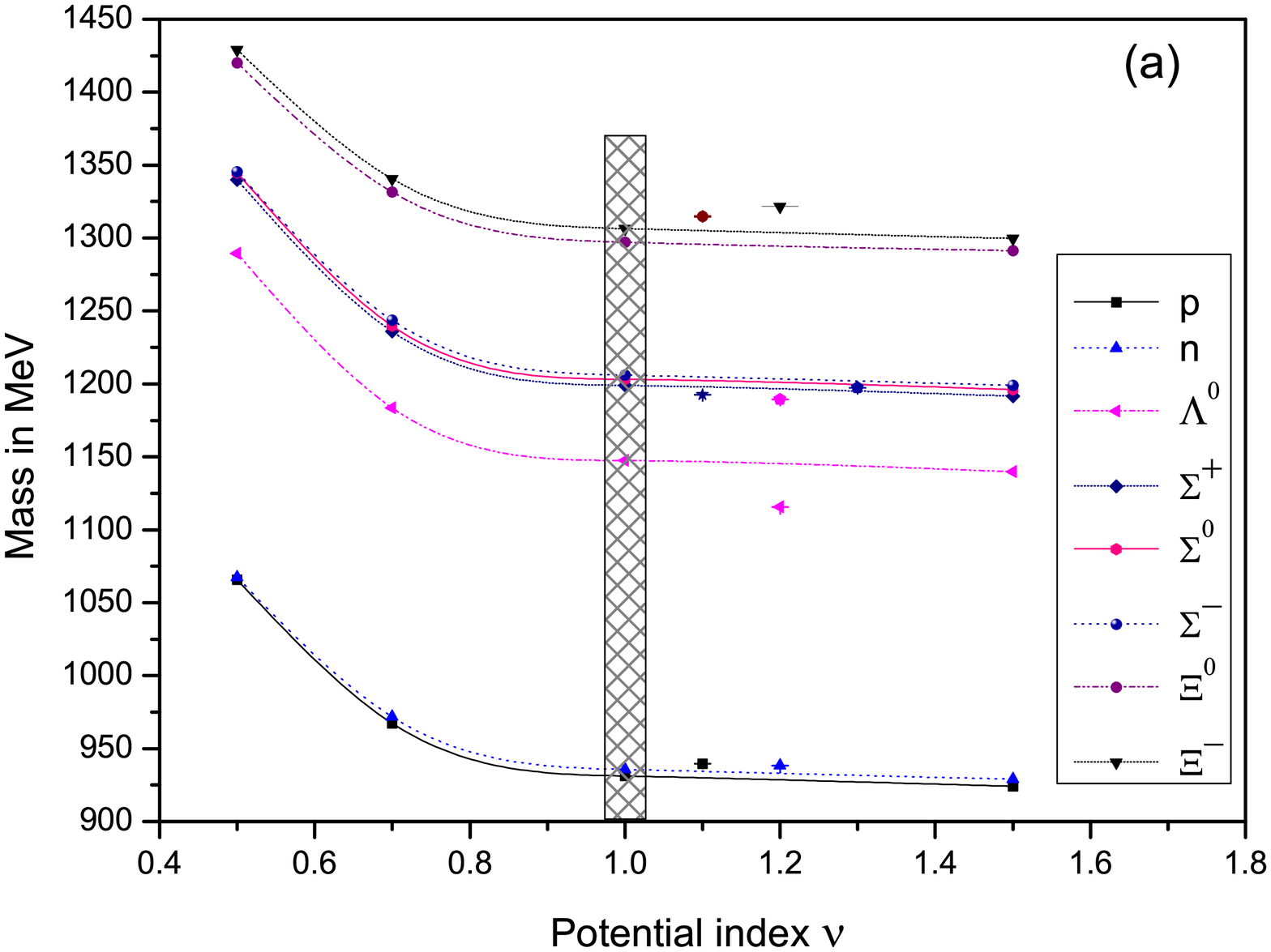}
\caption{Variation of octet  baryon masses with respect to potential
index $\nu$. Experimental masses are shown with
error bar. The shaded region show minimum root mean square deviation
with experimental results.} \label{fig:01}
\end{center}
\end{figure}

\begin{figure}
\begin{center}
\includegraphics[height=3.5in,width=3.75in]{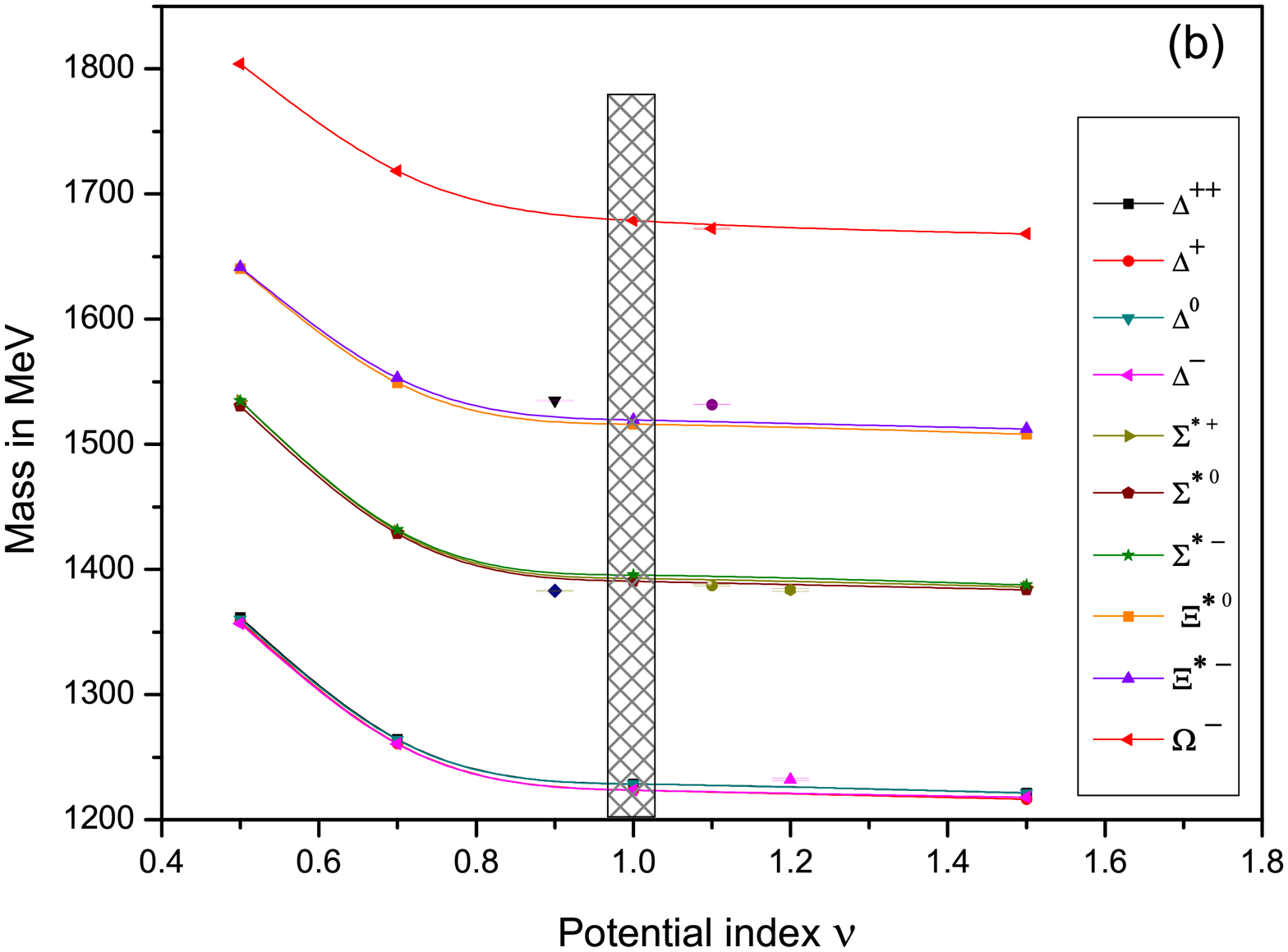}
\caption{Variation of  decuplet  baryon masses with respect to
potential index $\nu$. Experimental masses are
shown with error bar. The shaded region show minimum root mean
square deviation with experimental results.} \label{fig:02}
\end{center}
\end{figure}

\begin{figure}
\begin{center}
\includegraphics[height=3.5in,width=3.75in]{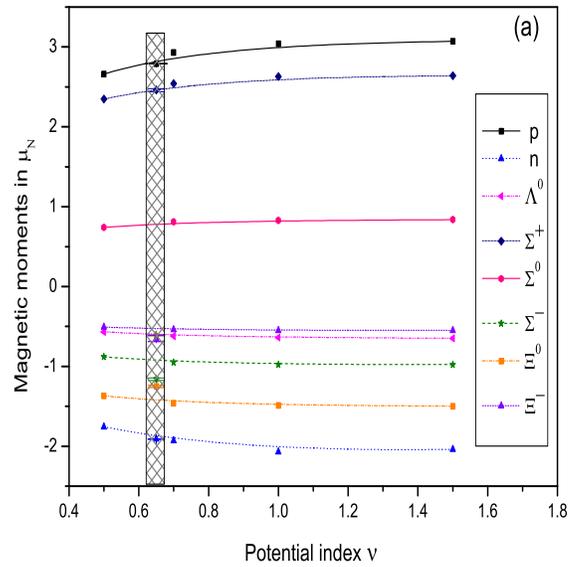}
\caption{Behaviour of the magnetic moments of octet baryons with
respect to potential index $\nu$. The known experimental values are
shown with error bar. The shaded region show minimum root mean
square deviation with experimental results.} \label{fig:03}
\end{center}
\end{figure}

\begin{figure}
\begin{center}
\includegraphics[height=3.5in,width=3.75in]{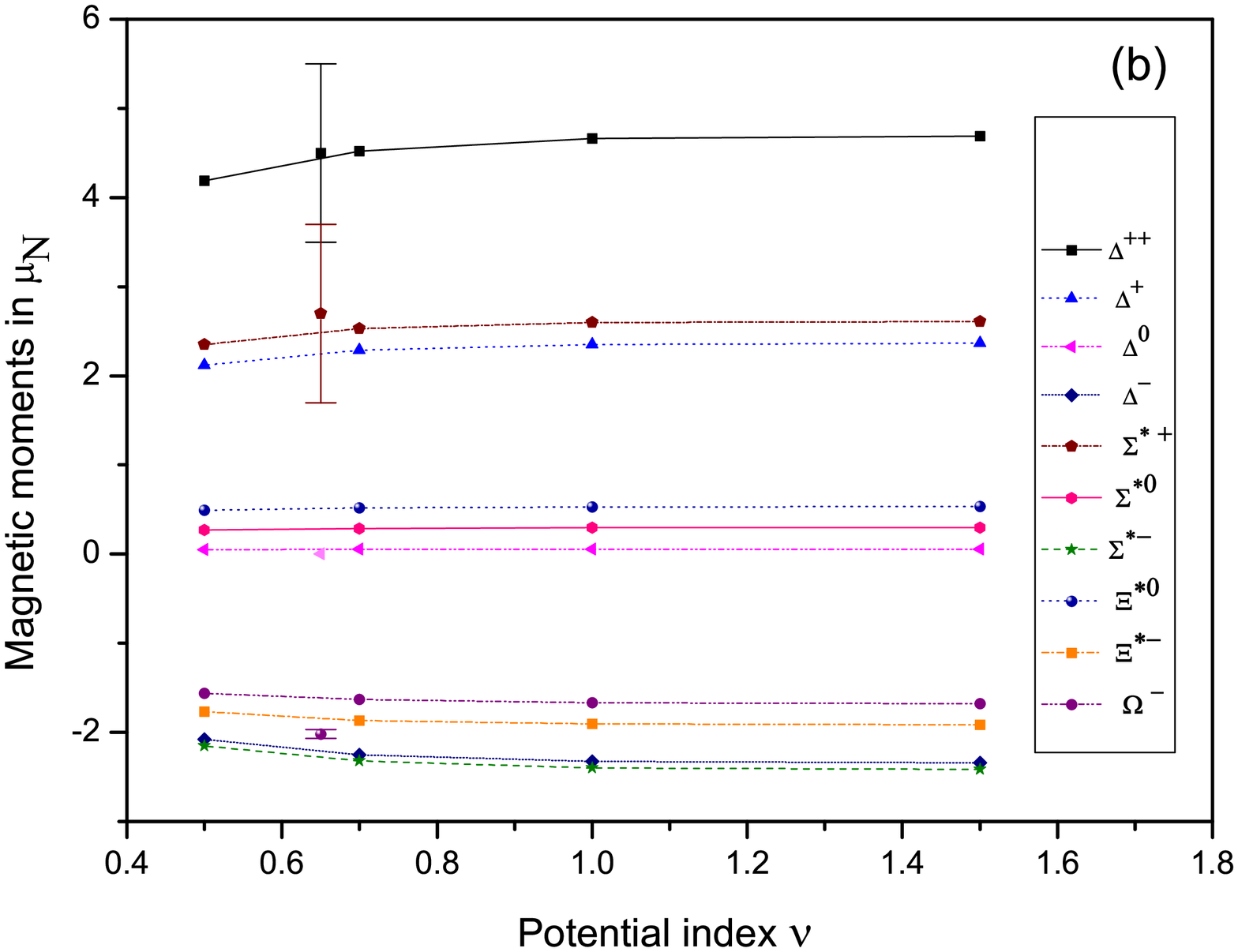}
\caption{Behaviour of the magnetic moments of  decuplet baryons with
respect to potential index $\nu$. The known experimental values are
shown with error bar.} \label{fig:04}
\end{center}
\end{figure}

\begin{table}[h]
\begin{center}
\caption{Masses of octet baryons ($J^P=\frac{1}{2}^+$)}\label{tab:1}
\begin{tabular}{lllllll}
\hline\hline

 hCPP$_\nu$&&\multicolumn{2}{c}{\textbf{\underline{Octet
Mass(MeV)}}} &&\multicolumn{2}{c}{\textbf{\underline{ Octet
Mass(MeV)}}}\\
Model & Baryon &Our&Others&Baryon&Our&Others\\
\hline
 0.5 &  uud(p)&   1065.68 &   939.00\cite{M.Bagchi 2006}   &uds($\Sigma^{0}$)& 1344.67 &  1193.00\cite{M.Bagchi 2006} \\
 0.7 &        &   967.41  &   938.27\cite{Phuoc Ha 2008}   &                 & 1239.94 &  1192.64\cite{Phuoc Ha 2008} \\
 1.0 &        &   931.08  &   866.00\cite{de Fisica 1986 } &                 & 1203.29 &  1022.00\cite{de Fisica 1986 }\\
 1.5 &        &   924.27  &   938.27\cite{PDG2008}         &                 & 1195.98 &  1192.64\cite{PDG2008}       \\
\hline

   0.5        &   ddu(n) &   1067.24 & 939.00\cite{M.Bagchi 2006}  &dds($\Sigma^{-}$)&     1345.46 &1197.00\cite{M.Bagchi 2006}   \\
   0.7              &    &   971.74  &   939.57\cite{Phuoc Ha 2008}  &     & 1243.71 & 1197.45\cite{Phuoc Ha 2008} \\
   1.0              &    &   935.77  & 866.00\cite{de Fisica 1986 }  & &  1205.99 &  1022.00\cite{de Fisica 1986 }  \\
   1.5              &    &   929.04  & 939.56\cite{PDG2008}      &   &    1199.06 &  1197.45\cite{PDG2008}         \\
\hline
0.5   &  uds($\Lambda^0$)&   1289.26 & 1116.00\cite{M.Bagchi 2006} &ssu($\Xi^{0}$)&   1420.19 & 1315.00\cite{M.Bagchi 2006}\\
0.7                 &    &   1183.59 &   1115.68\cite{Phuoc Ha 2008}  &  &  1331.65 &   1314.64\cite{Phuoc Ha 2008}      \\
1.0                 &    &   1147.34 &   1022.00\cite{de Fisica 1986 }    &     &  1297.09 & 1215.00\cite{de Fisica 1986 }         \\
1.5                 &    &   1139.88 & 1115.68\cite{PDG2008}      &     & 1291.43 & 1314.86\cite{PDG2008}           \\
\hline

0.5  &uus($\Sigma^{+}$) &   1339.95 & 1189.00\cite{M.Bagchi 2006}&ssd($\Xi^{-}$)&   1428.97 &1321.00\cite{M.Bagchi 2006}\\
0.7                 &    &   1235.98 &   1189.39\cite{Phuoc Ha 2008}    & &   1340.44 &   1321.39\cite{Phuoc Ha 2008}  \\
1.0                 &    &   1198.84 & 1022.00\cite{de Fisica 1986 }   & & 1306.55 &   1215.00\cite{de Fisica 1986 }   \\
1.5                 &    &   1191.50 & 1189.37\cite{PDG2008}      & &  1299.61 & 1321.71\cite{PDG2008}  \\
\hline\hline
\end{tabular}
\end{center}
\end{table}

\begin{table}
\begin{center}
\caption{Masses of decuplet baryons ($J^P=\frac{3}{2}^+$)}\label{tab:2}
\begin{tabular}{llllllll}
\hline\hline
 hCPP$_{\nu}$&&\multicolumn{2}{c}{\textbf{\underline{
Decuplet Mass(MeV)}}} &&\multicolumn{2}{c}{\textbf{\underline{
Decuplet
Mass(MeV)}}}\\
Model & Baryon  &Our&Others&Baryon&Our&Others\\
\hline
 0.5  &  uuu($\Delta^{++}$) & 1361.68 & 1232.00\cite{M.Bagchi 2006} &uds($\Sigma^{*0}$)& 1530.40 & 1384.00\cite{M.Bagchi 2006}\\
 0.7  &  &  1264.17 & 1230.82\cite{Phuoc Ha 2008}  & &   1428.43   &   1384.18\cite{Phuoc Ha 2008} \\
 1.0  & &   1228.63 & 1344.00\cite{de Fisica 1986 }  & &  1390.47   &  1447.00\cite{de Fisica 1986 }  \\
 1.5  &&   1221.21 & 1232.00\cite{PDG2008} &  &   1383.66   & 1383.70\cite{PDG2008} \\
\hline
 0.5&uud($\Delta^{+}$)& 1358.3 & 1232.00\cite{M.Bagchi 2006} &dds($\Sigma^{*-}$)& 1534.72 & 1387.00\cite{M.Bagchi 2006} \\
 0.7&                &1260.78 &1230.57\cite{Phuoc Ha 2008} & &  1431.66 &   1387.18\cite{Phuoc Ha 2008}  \\
 1.0&                &    1223.74   &   1344.00\cite{de Fisica 1986 } & &  1395.44   &   1447.00\cite{de Fisica 1986 }    \\
 1.5&               &   1216.33    &     1232.00\cite{PDG2008}    & &      1387.45  &   1387.20\cite{PDG2008} \\
\hline
0.5 &ddu($\Delta^{0}$)  &1360.22    &1232.00\cite{M.Bagchi 2006} &ssu($\Xi^{*0}$) & 1640.49 & 1532.00\cite{M.Bagchi 2006}    \\
0.7               &     & 1263.77& 1231.87\cite{Phuoc Ha 2008}   & &   1549.05  &   1531.81\cite{Phuoc Ha 2008} \\
1.0               & &  1228.69& 1344.00\cite{de Fisica 1986 } &   &   1516.19  &   1583.00\cite{de Fisica 1986 } \\
1.5               &   &   1221.25 & 1232.00\cite{PDG2008}      &  &   1508.03  &   1531.80\cite{PDG2008}     \\
\hline

0.5&ddd($\Delta^{-}$)    &   1356.79 & 1232.00\cite{M.Bagchi 2006}  &ssd($\Xi^{*-}$) & 1641.69 & 1535.00\cite{M.Bagchi 2006}       \\
0.7                      &&   1260.32 & 1234.73\cite{Phuoc Ha 2008}   &  &   1553.1  &   1534.95\cite{Phuoc Ha 2008}     \\
1.0                    & &   1223.65 & 1344.00\cite{de Fisica 1986 }  &  &   1519.35   &  1583.00\cite{de Fisica 1986 }     \\
1.5                    & &   1217.80 & 1232.00\cite{PDG2008}   &    &   1512.23  &  1535.00\cite{PDG2008}      \\
\hline

0.5&uus($\Sigma^{*+}$)& 1534.60& 1383.00\cite{M.Bagchi 2006}  &sss($\Omega^{-}$)&   1804.12 & 1672.00\cite{M.Bagchi 2006}  \\
0.7                    & &   1430.54   &   1382.74\cite{Phuoc Ha 2008} &&   1718.22 & 1672.45\cite{Phuoc Ha 2008}     \\
1.0                & & 1392.93      &   1447.00\cite{de Fisica 1986 }  &&   1678.7 & 1701.00\cite{de Fisica 1986 }   \\
1.5                    & &  1386.16       &   1382.80\cite{PDG2008}  &&  1668.16 &  1672.45\cite{PDG2008}  \\
\hline\hline
\end{tabular}
\end{center}

\end{table}

\begin{table}
\begin{center}
\caption{Magnetic moments of octet baryons ( in $\mu_N$) }\label{tab:3}

 \begin{tabular}{cccccccccc}
\hline\hline Various models& p & n & $\Lambda$ &
$\Sigma^+$&$\Sigma^0$&$\Sigma^-$&$\Xi^0$&$\Xi^-$\\
\hline
hCPP$_{\nu}$\\ $\nu=1.5$&3.07 &-2.04&-0.65&2.64&0.84&-0.98&-1.50&-0.55\\

 $\nu=1.0$  &3.04&-2.07&-0.64&2.63&0.83&-0.98&-1.49&-0.55\\

 $\nu=0.7$  &  2.93  & -1.93  & -0.62  & 2.54  & 0.81 &-0.95  & -1.46   &-0.54\\

$\nu=0.5$  &2.66&-1.76&-0.57&2.35&0.74&-0.88&-1.37&-0.51\\
\hline
EXPT. \cite{PDG2008}           &  2.79 &  -1.91 &  -0.61 & 2.46  &       &-1.16  & -1.25  & -0.65\\
\hline
QCDSR \cite{Lai Wang 2008}     &   2.82  &  -1.97 &  -0.56  & 2.31  & 0.69      &-1.16  & -1.15  & -0.64\\

$\chi$CQM \cite{H Dahiya 2009}& 2.80&-2.11&-0.66  & 2.39 & 0.54& -1.32& -1.24&-0.50\\

$\chi$PT \cite{L. S. Geng}& 2.58&-2.10&-0.66  & 2.43 & 0.66& -1.10& -1.27&-0.95\\

Latt \cite{D. B. Leinweber 1992}    & 2.79   & -1.60  &  -0.50  & 2.37  &  0.65 & -1.08  & -1.17  & -0.51\\

CDM \cite{Bae M. 1996}         & 2.79  &  -2.07   &-0.71 & 2.47  &       & -1.01  & -1.52  & -0.61\\

QM \cite{Gupta 1987}           & 2.79  & -1.91&  -0.59&  2.67&   0.78& -1.10&  -1.41&  -0.47\\

QM+T \cite{Gupta 1987}         &  2.79  & -1.91&  -0.61&  2.39&   0.63& -1.12&  -1.24&  -0.69\\

BAGCHI \cite{M.Bagchi 2006}    &  2.88&    -1.91&   -0.71&   2.59&    0.83&    -0.92 &  -1.45& -0.62\\

Dai fit A \cite{Dai J 1996}    &  2.84 &   -1.87&         &  2.46 &        &  -1.06 &  -1.28 & -0.61\\

Dai fit B \cite{Dai J 1996}    &  2.80  &   -1.92 &        &  2.46  &        &  -1.23 &  -1.26& -0.63\\

SIMON \cite{B. O. Kerbikov 2000}&2.54 &   -1.69 &  -0.69 &  2.48  &  0.80  &  -0.90 &   -1.49 & -0.63\\

SU(3)BR. \cite{J. W. Bos 1997}  &2.79 &  -1.97 & -0.60 & 2.48 &  0.66 &   -1.16&  -1.27& -0.65\\

PQM \cite{J. Franklin}& 2.68&-1.99&-0.56&2.52&&-1.17&-1.27&-0.59\\
\hline\hline
\end{tabular}
\end{center}
\end{table}

\begin{table}
\begin{center}
\caption{Magnetic moments of decuplet baryons (in $\mu_N$)}
\label{tab:4}
\begin{tabular}{@{}ccccccccccc}
\hline\hline
 Various\\ models&
 $\Delta^{++}$& $\Delta^{+}$& $\Delta^{0}$&
 $\Delta^{-}$&$\Sigma^{*+}$&$\Sigma^{*0}$&
 $\Sigma^{*-}$&
 $\Xi^{*0}$&$\Xi^{*-}$&$\Omega^-$\\
 \hline
 hCPP$_{\nu}$\\ $\nu=1.5$  & 4.69  &  2.37  & 0.05 & -2.34  & 2.61 & 0.29 & -2.42  & 0.53 &  -1.92&-1.68  \\
$\nu=1.0$ & 4.66  &  2.35  & 0.05 & -2.33  & 2.60  & 0.28 & -2.40  & 0.53 &  -1.91&-1.67  \\
$\nu=0.7$ & 4.52  &  2.29  & 0.05 & -2.25  & 2.53 & 0.27 & -2.32  & 0.52 &  -1.87&-1.63  \\
$\nu=0.5$ & 4.19  & 2.12   & 0.05 & -2.08  & 2.35 & 0.26 & -2.15  & 0.49 & -1.77&-1.56   \\
 \hline
 Expt.            & 4.5{$\pm$}0.95  &  2.70$^{+1.0}_{-1.3}$  &$\approx$0\  &        &      &      &        &        & &-2.02$\pm$0.06 \\
 \cite{PDG2008,M. Kotulla,{A. Bosshard 1991}}
 & 3.5-7.5\\
\hline
 LCQCD \cite{T. M. Aliev 2000}     & 4.40  &  2.20  & 0.00 & -2.20  & 2.70 & 0.20 & -2.28  & 0.40   &  -2.00&-1.56  \\

 QCDSR \cite{Frank X. Lee 1998}     & 4.39  &  2.19  & 0.00 & -2.19  & 2.13 & 0.32 & -1.66  & -0.69  &  -1.51&-1.49  \\

 Latt \cite{D. B. Leinweber 1992}     & 4.91  &  2.46  & 0.00 & -2.46  & 2.55 & 0.27 &  2.02  & 0.46   &  -1.68&-1.40  \\

$\chi$PT \cite{L. S. Geng}   &6.04 &2.84 & -0.36& -3.56  &3.07 & 0.00& -3.07   & 0.36 &  -2.56& -2.02       \\

$\chi$PT \cite{M. N. Butler 1994}& 4.00&2.10&-0.17  & -2.25& 2.00& -0.07& -2.20&0.10&-2.00&input\\

 RQM \cite{F. Schlumpf 1993}      & 4.76  &  2.38  & 0.00 & -2.38  & 1.82 & -0.27& -2.36  & -0.60  &  -2.41&-2.48  \\

 NRQM \cite{Ha P. and Durand L. 1990}       & 5.56  &  2.73  & -0.09& -2.92  & 3.09 & 0.27 & -2.56  & 0.63   &  -2.20&-1.81 \\

$\chi$QSM \cite{H. C. Kim 1998}  & 4.73  &  2.19  & -0.35& -2.9   & 2.52 & -0.08& -2.69  & 0.19   &  -2.48&-2.27  \\

$\chi$CQM \cite{H Dahiya 2009} & 4.51& 2.00 & -0.51& -3.02& 2.69&0.02 &-2.64&0.54&-1.84&-1.71\\

$\chi$B \cite{S. T. Hong 1994}    &3.59   &  0.75  & -2.09& -1.93  & 2.35 & -0.79& -3.87  & 0.58   &  -2.81&-1.75  \\

EMS \cite{Rohit Dhir 2009 }&
4.56&2.28&0.00&-2.28&2.56&0.23&-2.10&0.48&-1.90&-1.67\\

LCQCDSR\cite{K. Azizi}& 6.34 &3.17& 0.00&-3.17\\
\hline\hline
\end{tabular}
\end{center}
\end{table}

\begin{table}
\begin{center}
\caption{Magnitude of the transition Magnetic moments ($|\mu_{\frac{3}{2}^+\rightarrow \frac{1}{2}^+}|$) in
$\mu_N$} \label{tab:5}
\begin{tabular}{lcccccc}
\hline {\textbf{Decay Mode}}& \multicolumn{5}{c}{\textbf{\underline
{ Transition($|\frac{3}{2}^+\rightarrow \frac{1}{2}^+|$) Magnetic
moments($\mu_{N}$)}}}
 \\

&Expression&\multicolumn{2}{c}{\underline{$hCPP_\nu$}}&others& Expt. \cite{A. Bosshard 1991}\\
\hline

$\Delta^{+}\rightarrow p\gamma$&$|\frac{2\sqrt{2}}{3}(\mu_{u}-\mu_{d})|$&$\nu=0.5$&2.20&2.57 \cite{lang Yu} \\
&&$\nu=0.7$&2.40&2.76 \cite{S. T. Hong}\\
&&$\nu=1.0$&2.49&2.48 \cite{Rohit Dhir 2009 }\\
&&$\nu=1.5$&2.50&2.50 \cite{T. M. Aliev 2006}\\
\hline

$\Delta^{0}\rightarrow n\gamma$&$|-\frac{2\sqrt{2}}{3}(\mu_{d}-\mu_{u})|$&$\nu=0.5$&2.23&2.57 \cite{lang Yu} \\
&&$\nu=0.7$&2.42&2.76 \cite{S. T. Hong}& 3.23$\pm 0.1$\\
&&$\nu=1.0$&2.51&2.58 \cite{Rohit Dhir 2009 }\\
&&$\nu=1.5$&2.52&2.50 \cite{T. M. Aliev 2006}\\
\hline

$\Sigma^{*+}\rightarrow \Sigma^{+}\gamma$&$|\frac{2 \sqrt{2}}{3}(\mu_{u}-\mu_{s})|$&$\nu=0.5$&1.91&2.21 \cite{lang Yu} \\
&&$\nu=0.7$&2.06&2.24 \cite{S. T. Hong}\\
&&$\nu=1.0$&2.13&2.13 \cite{Rohit Dhir 2009 }\\
&&$\nu=1.5$&2.14&2.10 \cite{T. M. Aliev 2006}\\
\hline

$\Sigma^{*0}\rightarrow \Sigma^{0}\gamma$&$|\frac{\sqrt{2}}{3}(2\mu_{s}-\mu_{u}-\mu_{d})|$&$\nu=0.5$&0.89&0.88 \cite{lang Yu} \\
&&$\nu=0.7$&0.97&1.01 \cite{S. T. Hong}\\
&&$\nu=1.0$&1.00&0.96 \cite{Rohit Dhir 2009 }\\
&&$\nu=1.5$&1.01&0.89 \cite{T. M. Aliev 2006}\\
\hline

$\Sigma^{*0}\rightarrow \Lambda^{0}\gamma$&$|\frac{\sqrt{2}}{\sqrt3}(\mu_{u}-\mu_{d})|$&$\nu=0.5$&1.93&2.24 \cite{lang Yu} \\
&&$\nu=0.7$&2.09&2.46 \cite{S. T. Hong}\\
&&$\nu=1.0$&2.15&2.25 \cite{Rohit Dhir 2009 }\\
&&$\nu=1.5$&2.16&2.30 \cite{T. M. Aliev 2006}\\
\hline

$\Sigma^{*-}\rightarrow \Sigma^{-}\gamma$&$|\frac{2\sqrt{2}}{3}(\mu_{s}-\mu_{d})|$&$\nu=0.5$&0.21&0.44 \cite{lang Yu} \\
&&$\nu=0.7$&0.22&0.22 \cite{S. T. Hong}\\
&&$\nu=1.0$&0.23&0.22 \cite{Rohit Dhir 2009 }\\
&&$\nu=1.5$&0.23&0.31 \cite{T. M. Aliev 2006}\\
\hline

$\Xi^{*0}\rightarrow \Xi^{0}\gamma$&$|\frac{2\sqrt{2}}{3}(\mu_{u}-\mu_{s})|$&$\nu=0.5$&2.05&2.22 \cite{lang Yu}\\
&&$\nu=0.7$&2.17&2.46 \cite{S. T. Hong}\\
&&$\nu=1.0$&2.23&2.27 \cite{Rohit Dhir 2009 }\\
&&$\nu=1.5$&2.24&2.20 \cite{T. M. Aliev 2006}\\
\hline

$\Xi^{*-}\rightarrow \Xi^{-}\gamma$&$|\frac{2\sqrt{2}}{3}(2\mu_{s}-\mu_{d})|$&$\nu=0.5$&0.22&0.44 \cite{lang Yu}\\
&&$\nu=0.7$&0.23&0.27 \cite{S. T. Hong}\\
&&$\nu=1.0$&0.24&0.32 \cite{Rohit Dhir 2009 }\\
&&$\nu=1.5$&0.24&0.31 \cite{T. M. Aliev 2006}\\
\hline\hline
\end{tabular}

\end{center}
\end{table}

\begin{table}
\begin{center}
\caption{Radiative decay widths ($\Gamma_{R}$ in MeV) and branching ratio} \label{tab:6}
\begin{tabular}{llllclll}
\hline {\textbf{Decay
Mode}}&$hCPP_\nu$&\multicolumn{3}{c}{\textbf{\underline{Radiative
Decay Width($\Gamma_R$) in MeV  }}}
&\multicolumn{3}{c}{\textbf{\underline{Branching Ratio in $\%$}}} \\
&&Our&Others&Expt. &Symbol& our & Expt. \cite{PDG2008}\\
\hline

$\Delta^{+}\rightarrow p\gamma$&$\nu=0.5$&0.501&0.363 \cite{lang Yu}&&&0.424\\
&$\nu=0.7$&0.601&0.430 \cite{D. B. Leinweber}&&&0.510\\
&$\nu=1.0$&0.643&0.900 \cite{T. M. Aliev 2006}&0.64 \cite{S. Eidelman}&$\frac{\Gamma_R}{\Gamma(\Delta)}$&0.545&0.52-0.60\\
&$\nu=1.5$&0.644&&&&0.546\\
\hline

$\Delta^{0}\rightarrow n\gamma$&$\nu=0.5$&0.517&0.363 \cite{lang Yu}&&&0.438\\
&$\nu=0.7$&0.603&0.430 \cite{D. B. Leinweber}&&&0.511&\\
&$\nu=1.0$&0.655&0.900 \cite{T. M. Aliev 2006}&0.64 \cite{S. Eidelman}&$\frac{\Gamma_R}{\Gamma(\Delta)}$&0.555&0.52-0.60\\
&$\nu=1.5$&0.655&&&&0.555\\
\hline

$\Sigma^{*+}\rightarrow \Sigma^{+}\gamma$&$\nu=0.5$&0.111&0.100 \cite{lang Yu}&&&0.310\\
&$\nu=0.7$&0.129&0.100 \cite{D. B. Leinweber}&&&0.361\\
&$\nu=1.0$&0.137&0.11 \cite{T. M. Aliev 2006}&&$\frac{\Gamma_R}{\Gamma(\Sigma^{*+})}$&0.383&-\\
&$\nu=1.5$&0.139&&&&0.390\\
\hline

$\Sigma^{*0}\rightarrow \Sigma^{0}\gamma$&$\nu=0.5$&0.021&0.016 \cite{lang Yu}&&&0.058\\
&$\nu=0.7$&0.026&0.017 \cite{D. B. Leinweber}&&&0.072\\
&$\nu=1.0$&0.027&0.021 \cite{T. M. Aliev 2006}&$<$1.750 \cite{J. Colas}&$\frac{\Gamma_R}{\Gamma(\Sigma^{*0})}$&0.075\\
&$\nu=1.5$&0.027&0.022 \cite{E. Kaxiras}&&&0.077\\
\hline

$\Sigma^{*0}\rightarrow \Lambda^{0}\gamma$&$\nu=0.5$&0.216&0.241 \cite{lang Yu}&&&0.600\\
&$\nu=0.7$&0.265&&&&0.730\\
&$\nu=1.0$&0.274&0.470 \cite{T. M. Aliev 2006}&$<$2.100 \cite{T. S. Mast}&$\frac{\Gamma_R}{\Gamma(\Lambda^{*0})}$&0.763&1.3$\pm$0.4\\
&$\nu=1.5$&0.279&0.275 \cite{E. Kaxiras}&&&0.776\\
\hline

$\Sigma^{*-}\rightarrow \Sigma^{-}\gamma$&$\nu=0.5$&0.001&0.004 \cite{lang Yu}&&&0.003\\
&$\nu=0.7$&0.001&0.003 \cite{D. B. Leinweber}&&&0.003\\
&$\nu=1.0$&0.001&0.002 \cite{T. M. Aliev 2006}&$<$ 0.009 \cite{V. V. Molchanov}&$\frac{\Gamma_R}{\Gamma(\Sigma^{*-})}$&0.003&$<$ 0.024\\
&$\nu=1.5$&0.001&&&&0.003\\
\hline

$\Xi^{*0}\rightarrow \Xi^{0}\gamma$&$\nu=0.5$&0.185&0.131 \cite{lang Yu}&&&2.043\\
&$\nu=0.7$&0.200&0.129 \cite{D. B. Leinweber}&&&2.197\\
&$\nu=1.0$&0.216&0.140 \cite{T. M. Aliev 2006}&&$\frac{\Gamma_R}{\Gamma(\Xi^{*0})}$&2.378&$<$ 4.0\\
&$\nu=1.5$&0.211&&&&2.319\\
\hline

$\Xi^{*-}\rightarrow \Xi^{-}\gamma$&$\nu=0.5$&0.001&0.005 \cite{lang Yu}&&&0.019\\
&$\nu=0.7$&0.002&0.003 \cite{D. B. Leinweber}&&&0.021\\
&$\nu=1.0$&0.002&0.003 \cite{T. M. Aliev 2006}&&$\frac{\Gamma_R}{\Gamma(\Xi^{*-})}$&0.023&$<$ 4.0\\
&$\nu=1.5$&0.002&&&&0.023\\
\hline\hline
\end{tabular}

\end{center}
\end{table}

\begin{table}
\begin{center}
\caption{Percentage variation in the predictions of magnetic moments
of octet baryons} \label{tab:7}
\begin{tabular}{lccccccc}
\hline
system&$hCPP_\nu$&QCDSR&$\chi$CQM&$\chi$PT&BAGCHI&PQM&LATTICE\\
&(In shaded region)&\cite{Lai Wang 2008}&\cite{H Dahiya 2009}&\cite{L. S. Geng}&\cite{M.Bagchi 2006}&\cite{J. Franklin}&\cite{D. B. Leinweber 1992}\\
\hline
p&0.75&1.07&0.35&7.52&3.22&3.94&0.00\\
n&2.14&3.14&10.4&9.54&0.00&4.18&16.23\\
$\Lambda$&1.63&8.20&9.83&8.10&16.3&8.19&18.0\\
$\Sigma^+$&0.40&6.09&2.84&1.21&5.28&2.43&6.30\\
$\Sigma^-$&20.7&0.00&13.7&5.17&20.6&0.86& 6.80\\
$\Xi^0$&13.6&8.00&0.80&0.00&16.0&1.60&6.40 \\
$\Xi^-$&18.7&1.53&23.0&46.1&4.61&9.23&21.5\\
\hline
Average&8.20&4.00&8.70&11.00&9.43&4.34&10.74\\
\hline\hline
\end{tabular}

\end{center}
\end{table}

\section{Results and Discussion}
The masses of octet and decuplet baryons in the hypercentral coulomb
plus power potential ($hCPP_\nu$) model with the different choices
of potential index $\nu$ have been studied. Fig.(\ref{fig:01}) and
Fig.(\ref{fig:02}) show the behaviour of the predicted masses of the
octet and decuplet baryons with the potential index $\nu$ in the
range, 0.5 $\le \nu \le$ 1.5. The trend lines here show saturation
of the masses beyond $\nu$ $\ge$ 1.0. The shaded regions in
Fig.(\ref{fig:01}) and in Fig.(\ref{fig:02}) show the neighbour hood
region of $\nu$ at which the predicted masses are having
minimum root mean square deviation with the experimental masses.\\

The computed magnetic moments of the octet and decuplet baryons are
compared with the known experimental results as well as with other
model predictions in Table \ref{tab:3} and \ref{tab:4} respectively.
Present results for the choice of $\nu\approx$ 0.7  are found to be
in agreement with the known experimental values as well as with
other model predictions. Here, it should be noted that the better
agreement occur for the choice of $\nu$ ($0.6\le \nu \le0.7$)
slightly below the saturation region ($\nu \ge1$) (See Fig.\ref
{fig:01} - \ref{fig:04}). However the experimental measurements of
decuplet states are difficult and the known values for the $\Delta$
- baryons carry large errors \cite{PDG2008,M. Kotulla,{A.
Bosshard 1991}}.\\

The available experimental results for the $\Delta^{++}$ are
4.5$\pm$0.95 and 3.5-7.5 are in very good agreement with our
calculated magnetic moment 4.52 at $\nu$ = 0.7. The calculated
magnetic moments for $\Delta^{+}$, $\Delta^{0}$, and $\Omega^-$ are
also in good agreement with experimental result while comparing with
other theoretical models.\\

The behavior of the predicted magnetic moments of octet and decuplet
baryons with potential index $\nu$ are shown in Fig.(\ref{fig:03})
and Fig.(\ref{fig:04}) respectively. The same saturation trends
towards saturation beyond the potential index $\nu
>$ 1.0 are observed. The shaded region in Fig.(\ref{fig:03}) corresponds to
the region of $\nu$ $(0.6<\nu<0.7)$ for which the predicted octet
baryon magnetic moments show minimum root mean square deviation with
the experiments. The predicted magnetic moments of the decuplet
baryons in the same region of $\nu$ $(0.6<\nu<0.7)$ are found to be
closer to the existing experimental values of $\Delta$ and $\Omega$
baryons.\\

As the octet magnetic moments are known experimentally, we calculate
the percentage variations of the different model predictions with
respect to the experimental values and are given in Table
\ref{tab:7} for comparison. The present $hCPP_{\nu \approx 0.7}$
prediction for p, n, $\Lambda$, $\Sigma^+$ baryons are much better
with lesser percentage error compared to other model predictions.
And the average percentage variations from proton to $\Xi^-$
obtained from the Table \ref{tab:7} is about $8$ $\%$ only, while
that for the lattice predictions and that of $\chi$PT predictions is
about $10$ and $11$  $\%$ respectively. It can also be seen that the
predictions of QCDSR and PQM are having lower
variations of about 4  $\%$ only. \\

The transition magnetic moments obtained from the present study
($hCPP_\nu$ model) are in accordance with other theoretical
predictions with much less variations with the choices of $\nu$. However the experimentally known value for the transition magnetic
moments of (3.23$\pm 0.1$) $\Delta^{0} \rightarrow n\gamma$ is
higher than theoretical model predictions (see Table \ref{tab:5}).\\

The parameter free predictions of the radiative decay width for
$\Delta \rightarrow N\gamma$ (N=n,p) transitions obtained here are
in very good agreement with experiment compared to other model
predictions (see Table \ref{tab:6}). Prediction for other decuplet to
octet radiative transitions are well within the experimental
limits.\\

At the end, we like to point out important feature of the $hCPP_\nu$
model is the saturation behaviour of the predicted properties of the
baryons with $\nu > 1 $. Similar saturation behaviour was also
observed in the mass predictions of the $hCPP_\nu$ model in the
heavy flavour sector \cite{Bhavin 2008}.\\

It thus suggests that $hCPP_{\nu \ge1}$ model can adequately
represents the three body interactions among the quarks constituting
the baryons.\\

\textbf{Acknowledgement:} The authors acknowledge the financial
support from the University Grant commission, Government of India
under a Major research project F. 32-31/2006 (SR).\\


\section*{References}
\begin{thebibliography}{99}


\bibitem{PDG2008} Amsler C \emph{et al.} (Particle Data Group), Phys. Lett. \textbf{B 667}, 1 (2008).
\bibitem{M. Kotulla} M. Kotulla \emph{et al.} Phys. Rev. Lett.  \textbf{89}, 272001 (1991).
\bibitem{A. Bosshard 1991} A. Bosshard \emph{et al.}, Phys. Rev  \textbf{D 44}, 1962 (1991).
\bibitem{W. M. Yao} W. M. Yao \emph{et al.} (Particle Data Group),  J. Phys.  \textbf{G 33}, 1 (2006) and references therein.
\bibitem{D. B. Leinweber} D. B. Leinweber \emph{et al.} Phys. Rev  \textbf{D 48}, 2230 (1993).
\bibitem{E. Kaxiras} E. Kaxiras \emph{et al.} Phys. Rev  \textbf{D 32}, 695 (1985) and references therein.
\bibitem{S. Capstick} S. Capstick,  Phys. Rev  \textbf{D 46}, 1965 (1992).
\bibitem{M. N. Butler} M. N. Butler, M. J. Savage, and R. P. Springer, Nucl. Phys. \textbf{B 399}, 69 (1993).

\bibitem{T. M. Aliev 2006} T. M. Aliev and A. Ozpineci, Nucl. Phys. \textbf{B 732}, 291 (2006).


\bibitem{T. M. Aliev } T. M. Aliev, K. Azizi and A. Ozpineci, Phys. Rev  \textbf{D 79}, 056005 (2009).

\bibitem{M.Bagchi 2006}M. Bagchi, S. Daw, M. Dey, and J. Dey,  Europhys. Lett.,\textbf{75}, 548 (2006).
\bibitem{T. M. Aliev 2000} T. M. Aliev and A. Ozpineci, Phys. Rev  \textbf{D 62}, 053012 (2000).
\bibitem{Frank X. Lee 1998} Frank X. Lee, Phys. Rev  \textbf{D 57}, 1801 (1998).
\bibitem{D. B. Leinweber 1992}D. B. Leinweber, T. Draper, and R. M. Woloshyn , Phys. Rev. \textbf{D 46}, 3067 (1992).
\bibitem{I. C. Cloet 2003} I. C. Cloet, D. B. Leinweber and A. W. Thomas, Phys.
Lett. \textbf{B 563}, 157 (2003).
\bibitem{I. C. Cloet} I. C. Cloet, D. B. Leinweber and A. W. Thomas,
arXiv:nucl-th/0211027

\bibitem{L. S. Geng} L. S. Geng, J. Martin Camalich, and M. J. Vicente Vacas, Chinese Physics  \textbf{C 33}, X (2009); [arXiv:hep-ph/1001.0465].
\bibitem{M. N. Butler 1994}M. N. Butler, M. J. Savage, and R. P. Springer, Phys. Rev. \textbf{D 49}, 3459 (1994).
\bibitem{Meissner 1997} Meissner and S. Steininger, Nucl. Phys. \textbf{B 499}, 349 (1997).
\bibitem{P. Ha 1998} P. Ha and L. Durand, Phys. Rev. \textbf{D 58}, 093008 (1998).
\bibitem{S. J. Puglia 2000} S. J. Puglia and M. J. Ramsay, Phys. Rev. \textbf{D 62}, 034010 (2000).
\bibitem{F. Schlumpf 1993} F. Schlumpf, Phys. Rev. \textbf{D 48}, 4478 (1993).
\bibitem{K. T. Chao 1990} K. T. Chao, Phys. Rev. \textbf{D 41}, 920 (1990).
\bibitem{Ha P. and Durand L. 1990} Ha P. and Durand L., Phys. Rev. \textbf{D 58}, 093008 (1998).
\bibitem{Frank X. 1998} Frank X. Lee, Phys. Lett. \textbf{B 419}, 14 (1998).
\bibitem{Lai Wang 2008} Lai Wang and Frank X. Lee, Phys. Rev.  \textbf{D 78}, 013003 (2008).
\bibitem{H. C. Kim 1998} H. C. Kim, M. Praszalowicz, and K. Goeke,  Phys. Rev. \textbf{D 57}, 2859 (1998).
\bibitem{H. C. Kim 2004} H. C. Kim, M. Praszalowicz, Phys. Lett. \textbf{B 585}, 99 (2004).
\bibitem{H Dahiya 2009} Harleen Dahiya, Neetika Sharma and P. K. Chatley [arXiv:hep-ph/0912.5256v1].
\bibitem{S. T. Hong 1994} S. T. Hong and G. E. Brown, Nucl. Phys. \textbf{A 580},408 (1994).
\bibitem{M. I. Krivoruchenko 1987} M. I. Krivoruchenko, Sov. J., Nucl. Phys. \textbf{A 45},109 (1987).
\bibitem{Leinweber D. B. 1992} Leinweber D. B., Draper T. and Woloshyn R. M., Phys. Rev. \textbf{D 46}, 3067 (1992).


\bibitem{Bhavin 2008} Patel B, Rai A. K. and Vinodkumar P C,  J. Phys. G, \textbf{35}, 065001 (2008).
\bibitem{Bhavin J 2008} Patel B, Majethiya A and Vinodkumar P C, Pramana - J. Phys. {\bf 72}, 679 (2009).

\bibitem{Isgur1978} N.Isgur and G. Karl, Phys. Rev.  \textbf{D 18}, 4187 (1978);
 2653(1979); \textbf{D 20}, 1191 (1979); Phys. Lett. \textbf{B 72}, 109 (1977); Phys.
Lett. \textbf{B 74}, 353 (1978).
\bibitem{Godfrey1985} S. Godfrey et  al., Phys. Rev.  \textbf{D 32}, 189 (1985).
\bibitem {Capstick1986}S. Capstick and N. Isgur, Phys. Rev. \textbf{D 34}, 2809 (1986).
\bibitem {H. Dahiya 2003}H. Dahiya and M. Gupta, Phys. Rev. \textbf{D 67}, 114015 (2003); M. Gupta anf Navjot Kaur,Phys. Rev. \textbf{D 28}, 534 (1983); J. Singh and M. Gupta, J. Phys. \textbf{G 16}, L 45 (1990).
\bibitem {Murthy1985} M. V. N. Murthy, Z. Phys. C-Particles and Fields \textbf{31}, 81-86 (1986).
\bibitem {Roberts2007} W. Roberts and M. Pervin, [arXiv:hep-ph/0711.2492v1].

\bibitem{Garcilazo2007} H Garcilazo, J Vijande and A Valcarce, J. Phys. \textbf{G 34}, 961-976 (2007).

\bibitem {R. Bijker 2000} R. Bijker, F. Iachello and A. Leviatan, Annals of Physics \textbf{284},  89-133 (2000).
\bibitem {Santopinto1998} E. Santopinto, F. lachello and M. M. Giannini, Eur. Phys. J. A \textbf{1},  307- 315 (1998).
\bibitem{A 2008} Majethiya A, Patel B, and P C Vinodkumar, Eur. Phys. J \textbf{A 42}, 213 (2009).
\bibitem{Sameer2006} Sameer M. Ikhdair and Ramazan Sever, Int. Jn. Mod. Phys. \textbf{A 21}, 3989 (2006);
\textbf{18}, 4215 (2003); \textbf{19}, 1771 (2004); \textbf {21},
2190 (2006).
\bibitem{Heikkila1984} K. Heikkila, N. A. Tornquist and S. Ono, Phys. Rev.  \textbf{D 29}, 110 (1984).
\bibitem{Song1991} X. T. Song, J. Phys.  \textbf{G 17}, 49 (1991).
\bibitem{Mexicana 2004} J. G. Contreras, R. Huerta, and L. R. Quintero, Revista Mexicana de Fisica \textbf{50}, 490 (2004).
\bibitem{Rohit Dhir 2009 } Rohit Dhir and R. C. Verma, Eur. Phys. J \textbf{A 42}, 243 (2009).

\bibitem{Lang Yu} Lang Yu \emph{et al.} Phys. Rev. \textbf{D 66}, 033010 (2002) and references therein.
\bibitem{Phuoc Ha 2008} Phuoc Ha , J.Phys. \textbf{G 35}, 075006 (2008).
\bibitem{de Fisica 1986 } Mariaaline B. Do Vale \emph{et al.}, Revista brasileira de Fisica \textbf{16}, 4 (1986).

\bibitem{J. Franklin} J. Franklin, Phys. Rev. \textbf{D 73}, 114001 (2006).
\bibitem{Bae M. 1996} Bae M. and McGovern J. A., J. Phys.  \textbf{G 22}, 199 (1996).
\bibitem{Gupta 1987}S.K.Gupta and S.B.Khadkikar, Phys. Rev. \textbf{D 36}, 307 (1987).
\bibitem{Dai J 1996} Dai J., Dashen R., Jenkins E. and Manohar A. V., Phys. Rev.  \textbf{D 53}, 273 (1996).
\bibitem{B. O. Kerbikov 2000} B. O. Kerbikov and Yu. A. Simonov, Phys. Rev.  \textbf{D 62}, 093016 (2000).
\bibitem{J. W. Bos 1997} J. W. Bos \emph{et al.} Chinese J. of Phys. \textbf{35}, 2 (1997).


\bibitem{K. Azizi} K. Azizi, Eur. Phys. J \textbf{C 61}, 311 (2009).
\bibitem{lang Yu} Lang Yu \emph{et al.}, Phys. Rev.  \textbf{D 73}, 114001 (2006).

\bibitem{S. T. Hong} S. T. Hong, Phys. Rev.  \textbf{D 76}, 094029 (2007).
\bibitem{S. Eidelman} S. Eidelman \emph{et al.},  Phys. Lett. \textbf{B 592}, 1 (2004).
\bibitem{J. Colas} J. Colas \emph{et al.}, Nucl. Phys. \textbf{B 91}, 253 (1975).
\bibitem{T. S. Mast} T. S. Mast \emph{et al.}, Phys. Rev. Lett. \textbf{21}, 1715 (1968).
\bibitem{V. V. Molchanov} V. V. Molchanov \emph{et al.},  Phys. Lett. \textbf{B 590}, 161 (2004).
\end{thebibliography}
\end{document}